\documentclass{PoS}
\usepackage{multirow}
\usepackage{sidecap}
\usepackage{adjustbox}
\usepackage{fixltx2e}
\MakeRobust{\ref}

\newcommand\zeronubb {0\nu\beta\beta}

\newcommand\Eq[1]{Eq.~(\ref{eq:#1})}

\newcommand\Fig[1]{Fig.~\ref{fig:#1}}

\newcommand\Tab[1]{Table~\ref{tab:#1}}
\newcommand{\be}{\begin{equation}}
\newcommand{\ee}{\end{equation}}
\newcommand\beq{\begin{eqnarray}}
\newcommand\eeq{\end{eqnarray}}

\newcommand{\bfx}{{\mathbf x}}
\newcommand{\bfzero}{{\mathbf 0}}

\newcommand{\calR}{{\mathcal{ R}}}

\title{Neutrinoless double beta decay from lattice QCD}

\ShortTitle{Neutrinoless double beta decay from lattice QCD}

\author{\speaker{Amy Nicholson}%
         \thanks{LLNL-PROC-700398}\\
        Department of Physics, University of California, Berkeley\\
        E-mail: \email{anicholson@berkeley.edu}}

\author{Evan Berkowitz, Enrico Rinaldi, Pavlos Vranas\\
	Nuclear and Chemical Sciences Division, Lawrence Livermore National Laboratory\\
        E-mail: \email{berkowitz2@llnl.gov}, \email{rinaldi2@llnl.gov}, \email{vranas2@llnl.gov}}
     
\author{Chia Cheng Chang, Thorsten Kurth, Andr\'{e} Walker-Loud\\
Lawrence Berkeley National Laboratory\\
        E-mail: \email{chiachang@lbl.gov}, \email{tkurth@lbl.gov}, \email{awalker-loud@lbl.gov}}  
        
\author{M. A. Clark\\
NVIDIA Corporation\\
	E-mail: \email{mclark@nvidia.com}}        
          
\author{B\'{a}lint~Jo\'{o}\\
High Performance Computing Group, Thomas Jefferson National Accelerator Facility\\
        E-mail: \email{bjoo@jlab.org}}  
        
\author{Brian Tiburzi\\
Department of Physics, The City College of New York\\
RIKEN BNL Research Center, Brookhaven National Laboratory\\
	E-mail: \email{btiburzi@ccny.cuny.edu}}

\abstract{While the discovery of non-zero neutrino masses is one of the most important accomplishments by physicists in the past century, it is still unknown how and in what form these masses arise. Lepton number-violating neutrinoless double beta decay is a natural consequence of Majorana neutrinos and many BSM theories, and many experimental efforts are involved in the search for these processes. Understanding how neutrinoless double beta decay would manifest in nuclear environments is key for understanding any observed signals. In these proceedings we present an overview of a set of one- and two-body matrix elements relevant for experimental searches for neutrinoless double beta decay, describe the role of lattice QCD calculations, and present preliminary lattice QCD results.}

\FullConference{34th annual International Symposium on Lattice Field Theory\\
24-30 July 2016\\
University of Southampton, UK}

\begin{document}

\section{Introduction}

\begin{figure}
\begin{center}
%\beq
%\label{eq:tree}
\iffalse
\begin{fmffile}{simple13}
    \begin{fmfgraph*}(100,80)
        \fmfleft{i1,i2,i3,i4}
        \fmfright{o1,o2,o3,o4}
\fmf{plain}{i1,v1}
\fmflabel{d}{i1}
\fmflabel{d}{i4}
\fmflabel{u}{o1}
\fmflabel{u}{o4}
\fmflabel{e$^-$}{o2}
\fmflabel{e$^-$}{o3}
\fmf{plain}{v1,o1}
\fmf{plain}{o2,v2}
\fmf{plain}{v3,o3}
\fmf{phantom}{i2,v2}
\fmf{phantom}{i3,v3}
\fmf{plain}{i4,v4}
\fmf{plain}{v4,o4}
                   \fmf{photon,label=W$^-$,label.dist=3.4}{v1,v2}
                                      \fmf{plain,label=$\nu$,label.dist=4}{v3,v2}
                                      \fmf{photon,label=W$^-$,label.dist=3.4}{v4,v3}
    \end{fmfgraph*}
\end{fmffile}
%\eeq
\fi
\includegraphics[width=0.25\linewidth]{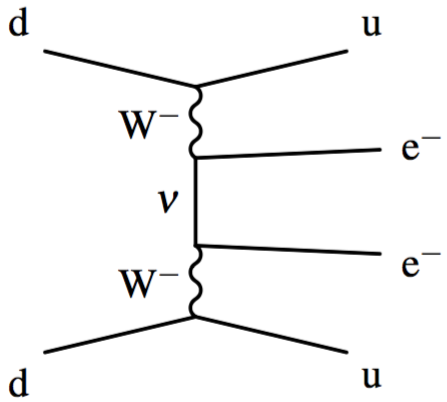}
\end{center}
\caption{\label{fig:neutrino}Quark-level Feynman diagram contributing to $\zeronubb$.}
\end{figure}

Neutrinoless double beta decay ($\zeronubb$) is a process that certain nuclei may undergo which has received a great deal of attention in recent years. Historically, $\zeronubb$ was proposed to occur shortly after beta decay was first understood. The story began in 1930, when Pauli first suggested the existence of the neutrino to accompany the electron in beta decay. Then, in 1932, Chadwick discovered the neutron, and both were incorporated into Fermi's effective theory of beta decay in 1934. Double beta decay was proposed to occur in some nuclei only one year later by Goppert-Mayer. Finally, in 1937 Majorana recognized that because the neutrino has no charge, it might be its own anti-particle, leading Racah to propose in that same year that a different type of beta decay, in which no neutrinos are emitted, could occur if Majorana's conjecture were true. This is because if the neutrino is its own anti-particle, then the same neutrino emitted from the first beta decay could be then absorbed to produce the second (see \Fig{neutrino}).
\\

This process violates lepton number by two units due to the production of two electrons without any corresponding anti-neutrinos. A look back at the history of lepton number reveals that it was initially introduced in an effort to explain why neutrinos produced in one type of reaction, such as the decay of a $\pi^{-}$ into a $\mu^{-}$ plus neutrino, always produce particles with the opposite charge, in our example a $\mu^{+}$, when incident upon a target. Furthermore, it was used to explain why $0\nu\beta\beta$ had not yet been observed in the relevant nuclei. Phase space calculations \cite{Haxton:1985am} imply that the neutrinoless mode should in fact be greatly favored over standard double beta decay if it is allowed. Therefore, the conclusion from these observations was that there must be some quantum number, deemed lepton number, which forbids the $0\nu\beta\beta$, and explains why neutrinos emitted in decays only produce one type of lepton when incident upon a target. 

It wasn't understood, however, until the 1950's that the charged current weak interaction is maximally parity violating, and therefore, neutrinos are always produced or absorbed with a specific helicity. With this discovery, lepton number was no longer needed to explain the experimental data, because helicity exactly forbids the second vertex in $0\nu\beta\beta$. Finally, near the turn of the millenium, experimentalists confirmed the existence of neutrino oscillations, proving that neutrinos are not massless, as the Standard Model presumes. This observation removes the helicity labels forbidding $0\nu\beta\beta$, because helicity is not conserved for massive particles. This means that there is a chance for a helicity flip, proportional to the mass of the neutrino, leading to an allowed absorption of the neutrino in a second beta decay. In this way, $0\nu\beta\beta$ experiments are sensitive to the absolute mass scale of the neutrino, which oscillation experiments cannot predict. However, since we know that the mass of the neutrino must be extremely tiny, at most a couple of eV, this process must be highly suppressed compared to the standard double beta decay. Because double beta decay in any form also involves a second-order weak interaction, the half-life for $0\nu\beta\beta$ is extremely long, with a current bound of about $10^{25}$ years \cite{Geesaman:2015fha}. 

While experiments have confirmed that neutrinos have mass, we do not yet know whether the mass is Majorana or Dirac in nature. Observation of $\zeronubb$ would unequivocally confirm that neutrinos have a Majorana mass component. There are good arguments for why we expect this to be the case. The simplest argument is that whatever is not forbidden should occur in nature. We know that lepton-number is not respected by the Standard Model due to anomalies. Non-zero neutrino masses are already an indication of beyond-the-Standard-Model physics, so if we accept that the Standard Model is an effective theory, then the first dimension-5 operator we can write down involving Standard Model fields and obeying Standard Model symmetries corresponds to a Majorana neutrino mass term, 
\beq
\label{eq:effMaj}
\mathcal{L}_5 = -\frac{c_5}{\Lambda_{\mathrm{UV}}} \left(\bar{L}\tilde{H} \right)\left(\tilde{H} L\right)^{\dagger} \ ,
\eeq
in which the neutrino mass is proportional to $m_\nu \propto c_5 v^2 / \Lambda_{\mathrm{UV}}$, where $v$ is the Higgs vacuum expectation value. Since the operator is allowed by symmetry, it would require an extreme amount of fine-tuning, or some other as-yet-unknown symmetry, to eliminate this mass term. 

Majorana masses are even more appealing because they may answer the question of why the observed neutrinos are so much lighter than other known Standard Model particles. If the neutrinos have only a Dirac mass, arising from coupling to the Higgs field in the usual way along with some right-handed neutrinos which have not yet been observed, then fine-tuning of the coupling to the Higgs field is required for the neutrinos to be much lighter than other Standard Model particles. A solution to this fine-tuning problem is given by the so-called seesaw mechanism. Since, again, everything that is not explicitly forbidden should occur, once we allow right-handed neutrinos to exist in our theory then we should write down the most general mass matrix for left- and right-handed neutrinos,
\begin{eqnarray}
\left(\begin{tabular}{cc}
$M_L$ & $M_D$ \\
$M_D$ & $M_R$
\end{tabular}\right) \ ,
\end{eqnarray}
where $M_L,M_R$ are the Majorana masses for the left- and right-handed neutrinos, respectively, $M_D$ is the Dirac mass, and this is a reduced block of the more general $4\times4$ mass matrix including both chiralities (and we are assuming only one generation of neutrino for simplicity). We should set $M_L=0$ because an explicit, dimension-3 Majorana mass operator is forbidden by the electroweak symmetry. Now, finding the eigenstates of this mass matrix leads us to two Majorana neutrinos with masses,
\begin{eqnarray}
m_l \sim M_D^2/M_R \qquad m_h \sim M_R \ ,
\end{eqnarray}
where we have assumed that $M_R \gg M_D$ since right-handed neutrinos have not yet been observed. Thus, the heavier the right-handed neutrino is, the lighter the observed neutrinos are. We can be somewhat quantitative in our estimate if we assume that the Dirac mass is of the scale of other Standard Model particles and set it equal to the top quark mass, and the light neutrinos have masses of order the mass splitting, $\Delta_{23}$, seen in oscillation experiments, then we have,
\begin{eqnarray}
M_D \sim 200 \mathrm{~GeV} \ , \qquad m_l \sim 0.05 \mathrm{~eV} \ , \qquad m_h \sim M_R \sim 10^{15} \mathrm{~GeV} \ ,
\end{eqnarray}
giving us a right-handed neutrino which exists near the Grand Unified Theory scale. 

Finally, it should be noted that right-handed neutrinos are not a necessary component of the seesaw mechanism. Integrating out the right-handed neutrinos leads to the effective Majorana mass operator, \Eq{effMaj}, with a coefficient suppressed by the scale of the heavy mass. Any other form of new physics occurring at a large mass scale that leads to a Majorana neutrino mass will also display a seesaw mechanism, with the light neutrino mass suppressed by the scale of the new physics.

Aside from confirming that neutrinos are Majorana in nature, observation of $\zeronubb$ could give us a source of lepton number violation that is much more significant than that produced by the Standard Model via anomalies. Such an observation could lead to enormous consequences for cosmology. As an example, if heavy right-handed neutrinos exist, then they would be present following the Big Bang, and might then undergo $CP$-violating decays to lighter leptons \cite{Pascoli:2006ci}. This is then a source for leptogenesis. Because the Standard Model conserves baryon number minus lepton number ($B-L$) exactly, then Standard Model processes, called sphalerons, can convert an excess of leptons in the early universe to an excess of baryons over anti-baryons. This gives us a source of baryogenesis, which for reasonable neutrino model parameters might be sufficient to explain the current excess of matter over anti-matter in the universe (see, e.g., \cite{Davidson:2008bu,Buchmuller:2005eh}, and references therein).

Experimentally, nuclear environments provide a natural filter for the double beta decay process. For example, when plotting the mass excess of the set of $A=76$ nuclei versus atomic number, $Z$, one finds empirically that the masses follow roughly two parabolas, one parabola falling through the masses of even $Z$ nuclei and a separate one, shifted toward higher mass excess, through nuclei having odd $Z$. The reason for the shift toward higher mass for odd $Z$ nuclei is nuclear pairing, in which neutrons and protons of opposite spin form pairs. When this occurs a non-zero amount of energy, the pairing gap, is required to break a pair. Nuclei having even $A$ and odd $Z$ contain one unpaired proton and one unpaired neutron, leading to an overall upward shift in energy per nucleon compared to adjacent even $Z$ nuclei with the same $A$. The consequence is that for certain nuclei, such as $^{76}$Ge, the cost in energy for breaking two pairs in order to undergo a single beta decay to $^{76}$As is too great, and is therefore energetically forbidden. On the other hand, double beta decay, in which two neutrons decay into two protons, is allowed, so that $^{76}$Ge may decay to $^{76}$Se. 

The two types of double beta decay, the neutrinoless and two-neutrino modes, may be differentiated from each other experimentally using spectroscopic methods. Because neutrinos carry away missing energy in the decay, the two-neutrino decay mode displays a broad energy profile for the resulting two electrons, with an upper bound at the total energy ($Q$-value) of the nuclear transition. The strength of the two-neutrino decay at the $Q$-value is essentially zero. However, for the neutrinoless mode, the two electrons must carry all of the energy of the transition, leading to a delta function in the energy profile at the $Q$-value, with some broadening due to detector resolution. There are large experimental efforts planned and underway across the globe looking for such signatures, such as the Cuore and GERDA experiments operating at Gran Sasso in Italy, using $^{130}$Te and $^{76}$Ge as sources, respectively, SNO$^+$ in Ontario, Canada, also using $^{130}$Te, and NEXO in New Mexico, USA, with a planned $^{136}$Xe source, to name a few (for a recent experimental review, see \cite{Dell'Oro:2016dbc}).

\section{Potential lattice QCD inputs}

The standard picture of $\zeronubb$ involves the long-range exchange of a light neutrino. At the microscopic level, the rate for this process depends on the well-known axial coupling of the nucleon, $g_A$. One potential contribution to better understanding of this type of decay in actual nuclei is through better understanding of so-called $g_A$ quenching, or in-medium modifications of the axial coupling. While $g_A$ quenching can be inferred to some extent from experimental data on single beta decay of other nuclei, the energy involved in $\zeronubb$ can be on the order of 100 MeV \cite{Bilenky:2012qi}, much larger than the energies involved in single beta decay. Therefore, calculations of the axial form factors for multi-nucleon systems may be useful. However, currently by far the largest uncertainty in such long-range transition amplitudes comes from ill-understood many-body effects such as truncation of the many-body wavefunction in nuclear models, with discrepancies between models on the order of $100\%$. Until these systematics are under better control, understanding corrections to $g_A$ due to quenching at non-zero momentum, which are expected to be moderate, may not lead to significant improvement in understanding this process in nuclei.

On the other hand, contributions from short-range operators are essentially unknown, so lattice QCD may be able to make the biggest impact through calculations of these matrix elements. Short-range operators are produced after integrating out heavy modes which might contribute to the $\zeronubb$ process. For example, the decay may occur due to the exchange of a heavy right-handed neutrino (which contributes to the seesaw mechanism discussed above) in a left-right symmetric extension of the Standard Model. While na\"ively one might expect that such processes will be suppressed due to the heavy particle propagator, which scales as $\sim 1/M_R$ for right-handed neutrinos, recall that the long-range, light neutrino exchange process requires a helicity flip, and will be proportional to the mass of the light neutrino, which we expect from the seesaw mechanism to also scale as $\sim 1/M_R$. Thus, whether the short- or long-range operators dominate depends on the details of the particular model under investigation. Furthermore, in order to differentiate between these types of operators and learn about the mechanism behind the $\zeronubb$ process, we must perform a quantitative comparison of the contributions from each operator. 

It should be noted that no matter what the mechanism behind $\zeronubb$, observation of this process always indicates that neutrinos are Majorana particles, because we may rearrange any $\zeronubb$ diagram to form a black box Majorana mass diagram \cite{schechter1982neutrinoless,Nieves:1984sn,Takasugi:1984xr,Rosen:1992qa,Hirsch:2006yk}. The decay may not even involve neutrinos at all. For example, R-parity violating supersymmetric interactions involving the exchange of charged leptonic superpartners can also lead to $\zeronubb$. Thus, $\zeronubb$ experiments may be used to impose constraints on R-parity violating couplings in supersymmetric models, which are important for understanding stability of the lightest superpartner, a dark matter candidate. However, in order to make connections between experimental signatures and models such as supersymmetry, we must first calculate the contributions from the relevant short-range operators.

\section{Effective operators}
\begin{figure}
\includegraphics[width=\linewidth]{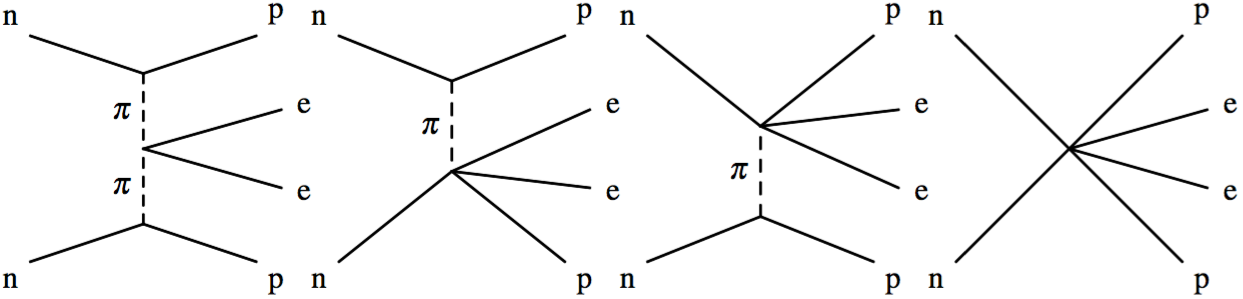}
\caption{\label{fig:EFTOps}Tree-level diagrams involving short-range operators contributing to $\zeronubb$ in the chiral effective theory.}
\end{figure}

We may use chiral effective theory to categorize a set of possible hadronic interactions arising from short-ranged operators that contribute to $\zeronubb$ (\Fig{EFTOps}) \cite{Prezeau:2003xn}. There is a long-range contribution coming from the exchange of a pion between two nucleons. The $\zeronubb$ occurs as a contact operator converting the $\pi^-$ to a $\pi^+$. In addition, there are contact operators involving one nucleon plus a pion, as well as a two-nucleon contact operator. The first diagram, corresponding to long-range pion exchange, is enhanced by the two pion propagators and is therefore leading order, while the two-nucleon plus one pion contact operator contributes at next-to-leading order. Finally, the two-nucleon contact operator, containing no light pion propagators, is next-to-next-to-leading order. 
\\

We will focus on the calculation of the leading order, $\pi^-$ to $\pi^+$ contact operator that appears in the left-most panel of \Fig{EFTOps}. In the context of effective field theory, what we calculate is the low-energy constant corresponding to the matrix element of this contact operator acting on a single on-shell pion. Once this low-energy constant has been evaluated, chiral effective field theory then tells us the rate of the $\zeronubb$ occurring between two nucleons via pion exchange.

There are in general nine four-quark contact operators appearing in the effective Lagrangian contributing to short-ranged $\zeronubb$, as outlined in \cite{Prezeau:2003xn}. For the $\pi^- \to \pi^+$ transition, we only require parity even operators. Furthermore, contributions from vector operators may be shown to be suppressed by the electron mass \cite{Prezeau:2003xn}. Therefore, we calculate the following four-quark operators:
\beq
\label{eq:Ops}
\mathcal{O}_{1+}^{++} &=& \left(\bar{q}_L \tau^+ \gamma^{\mu}q_L\right)\left[\bar{q}_R \tau^+\gamma_{\mu} q_R \right] \ , \cr
\mathcal{O}_{2+}^{++} &=& \left(\bar{q}_R \tau^+ q_L\right)\left[\bar{q}_R \tau^+ q_L \right] + \left(\bar{q}_L \tau^+ q_R\right)\left[\bar{q}_L \tau^+ q_R \right] \ , \cr
\mathcal{O}_{3+}^{++} &=& \left(\bar{q}_L \tau^+ \gamma^{\mu}q_L\right)\left[\bar{q}_L \tau^+ \gamma_{\mu} q_L \right] + \left(\bar{q}_R \tau^+ \gamma^{\mu}q_R\right)\left[\bar{q}_R \tau^+ \gamma_{\mu} q_R \right] \ , 
\eeq
where the Takahashi notation $()$ or $[]$ denotes color indices which are contracted together \cite{Takahashi:2012}. 
In addition, we calculate the following color-mixed operators:
\beq
\mathcal{O}_{1+}^{'++} &=& \left(\bar{q}_L \tau^+ \gamma^{\mu}q_L\right]\left[\bar{q}_R \tau^+\gamma_{\mu} q_R \right) \ , \cr
\mathcal{O}_{2+}^{'++} &=& \left(\bar{q}_L \tau^+ \gamma^{\mu}q_L\right]\left[\bar{q}_L \tau^+ \gamma_{\mu} q_L \right) + \left(\bar{q}_R \tau^+ \gamma^{\mu}q_R\right]\left[\bar{q}_R \tau^+ \gamma_{\mu} q_R \right) \ , 
\eeq
which arise at the QCD scale through renormalization of the weak-scale operators and will mix with the operators of \Eq{Ops} \cite{Graesser:2016bpz}. Note that the $\Delta I=2$ operator basis is essentially the same as the $\Delta F=2$ basis needed for the calculation of $B_K$ in general beyond-the-standard model scenarios \cite{Buras:2000if}.

This set of operators may be organized by the order in which they appear in chiral effective theory. As discussed in \cite{Prezeau:2003xn}, $\mathcal{O}_{3+}^{++}$ does not contain a leading order contribution, and is therefore expected to scale with $m_{\pi}^2$. Furthermore, this operator can be related by effective theory to $K\to\pi\pi$ decay \cite{Savage:1998yh}, so there may be further suppression as part of the $\Delta I =1/2$ rule. In addition to the ordering according to chiral counting, the sizes of the contributions from these operators depends heavily on the particular model under consideration. For example, in left-right symmetric models with no mixing between the left- and right-handed $W$ bosons, $\mathcal{O}_{1+}^{++}$ and $\mathcal{O}_{2+}^{++}$, which mix left- and right-handed currents, vanish, leaving only $\mathcal{O}_{3+}^{++}$ to contribute. For models with mixing, the relative sizes of the different operators depend on the strength of the mixing and the masses of the $W$ boson eigenstates \cite{Prezeau:2003xn}.  

\section{Lattice calculation}
\begin{figure}
\begin{center}
\includegraphics[width=0.25\linewidth]{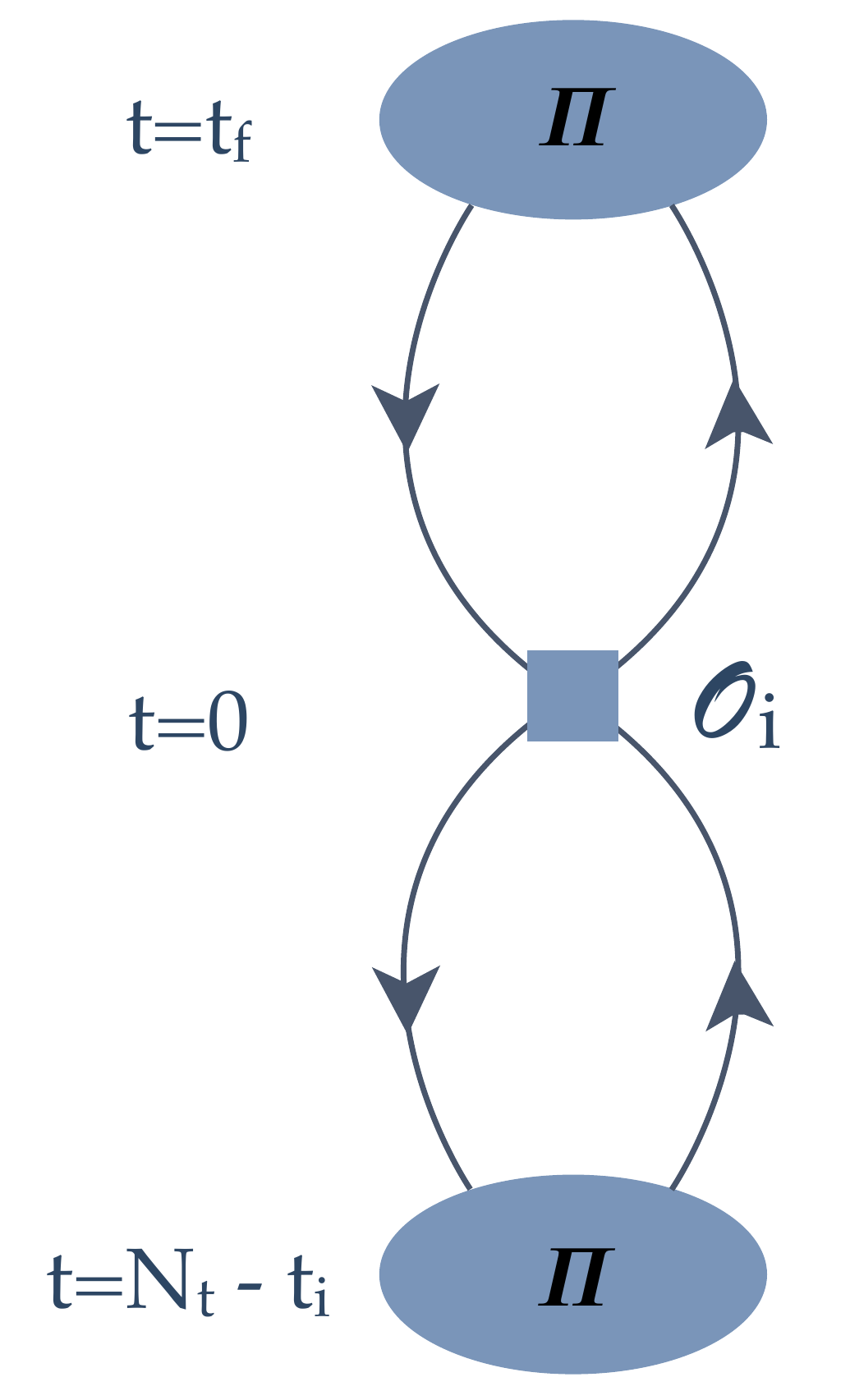}
\end{center}
\caption{\label{fig:Ops}Sketch of the contraction scheme used to calculate the $\pi \to \pi$ transition operators. Pion blocks, projected onto definite momenta, are created at times $N_t-t_i$ and $t_f$, where $N_t$ is the length of the temporal extent of the lattice. All quark propagators are contracted with the operator $\mathcal{O}_i$ at a single spacetime point.}
\end{figure}

The setup of the lattice calculation is as follows: we create a pion block,
\beq
\Pi_{a,\alpha,b,\beta} = \sum_{c,\gamma}\sum_{\bfx} \left[ S_{d}\left(\bfx,t;\bfzero,0\right) \gamma_5 \right]_{b,\beta,c,\gamma}\left[ S^{\dagger}_{u}\left(\bfx,t;\bfzero,0\right) \gamma_5 \right]_{a,\alpha,c,\gamma} \ ,
\eeq
which is fully contracted at one time, with open spin and color indices. These open indices are then tied up with an operator at a single spacetime point, with $t=0$. A second pion block is created at a later time and propagates backward toward the operator insertion (see \Fig{Ops}). This setup is similar in spirit to calculations of $K^0$-, $D^0$- and $B^0_{(s)}$-meson mixing (for a review, see \cite{Aoki:2016frl}), and $n\bar{n}$ oscillations \cite{Buchoff:2012bm}. 

Because all quark propagators are tied up with the operator (at a single lattice point), we are able to perform an exact momentum projection at both source and sink without having to calculate all-to-all propagators. For now we only perform calculations at zero total momentum and zero momentum transfer, but this setup is easily generalized for non-zero momenta.

We use the publicly available highly-improved staggered quark (HISQ) gauge field configurations produced by the MILC collaboration \cite{Bazavov:2012xda,Bazavov:2015yea}. This set of configurations includes a number of different lattice spacings, volumes, and pion masses, including physical pion mass, which may be used to analyze systematic effects and perform extrapolations. We have performed calculations on the ensembles tabulated in \Tab{ens}. 

\begin{table}
\label{tab:ens}
\begin{center}
\begin{tabular}{cccc}
$a (\mathrm{fm})$&$m_{\pi} \sim 310 \mathrm{~MeV}$ & $m_{\pi}  \sim 220 \mathrm{~MeV}$ & $m_{\pi}  \sim 135 \mathrm{~MeV}$ \\
\hline
0.15 & $16^3\times 48, m_{\pi} L \sim 3.78$ & $24^3\times 48, m_{\pi} L\sim 3.99$ & $32^3\times 48, m_{\pi} L\sim 3.25$ \\
0.12 & & $24^3\times 64, m_{\pi}L \sim 3.22$ & \\
0.12 &$24^3\times 64, m_{\pi} L\sim 4.54$&$32^3\times 64, m_{\pi}L \sim 4.29$&$48^3\times 64, m_{\pi} L\sim 3.91$ \\
0.12 & & $40^3\times 64, m_{\pi} L\sim 5.36$ & \\
0.09 &$32^3\times 96, m_{\pi} L\sim 4.50$ & $48^3\times 96, m_{\pi} L\sim 4.73$ & \\
\end{tabular}
\end{center}
\caption{List of HISQ ensembles used for this calculation, showing the volumes studied for a given lattice spacing and pion mass.}
\end{table}

On this set of ensembles, we have calculated M\"obius domain wall quark propagators \cite{Brower:2012vk,Action} using the irresponsibly fast solver in the QUDA library \cite{Clark:2009wm,Babich:2011np}. This mixed-action setup is beneficial because mixing between operators having different chiral symmetry is exponentially suppressed due to the better chiral symmetry respected by the valence propagators. We use the gradient flow method \cite{Luscher:2013cpa,Luscher:2011bx,Luscher:2010iy,Narayanan:2006rf} for smearing the gauge field configurations \cite{Action,GradFlow}, which we find gives good control over $m_{\mathrm{res}}$ at moderate $L_5$ with optimal values of $m_5 \leq 1.3$. Finally, we utilize both wall and point sources for our pion fields to aid in the assessment of excited state contamination. Currently, we have approximately 1000 sources for each ensemble.

\section{Results}

\begin{figure}
\begin{center}
\includegraphics[width=0.45\linewidth]{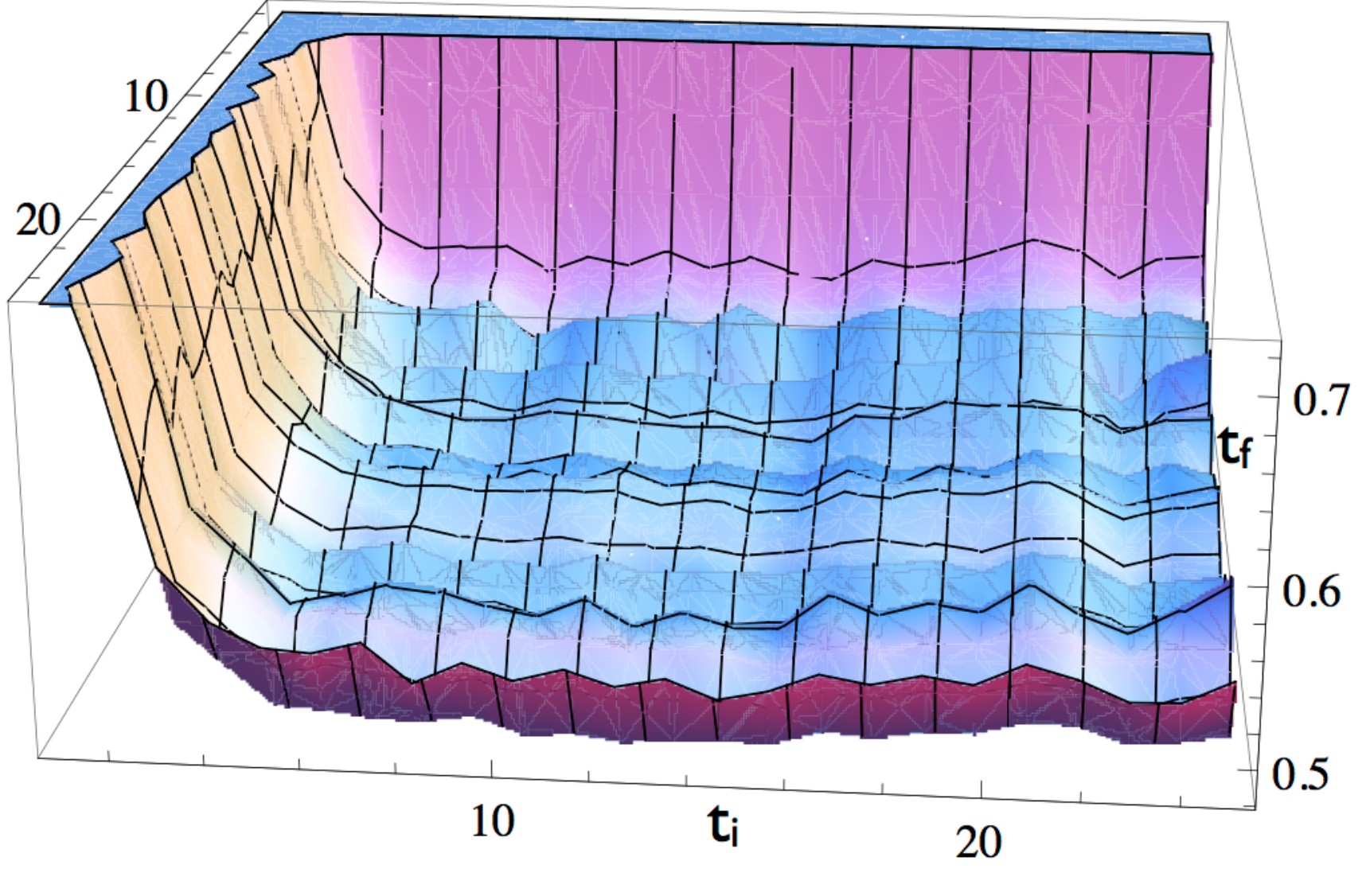}
\includegraphics[width=0.45\linewidth]{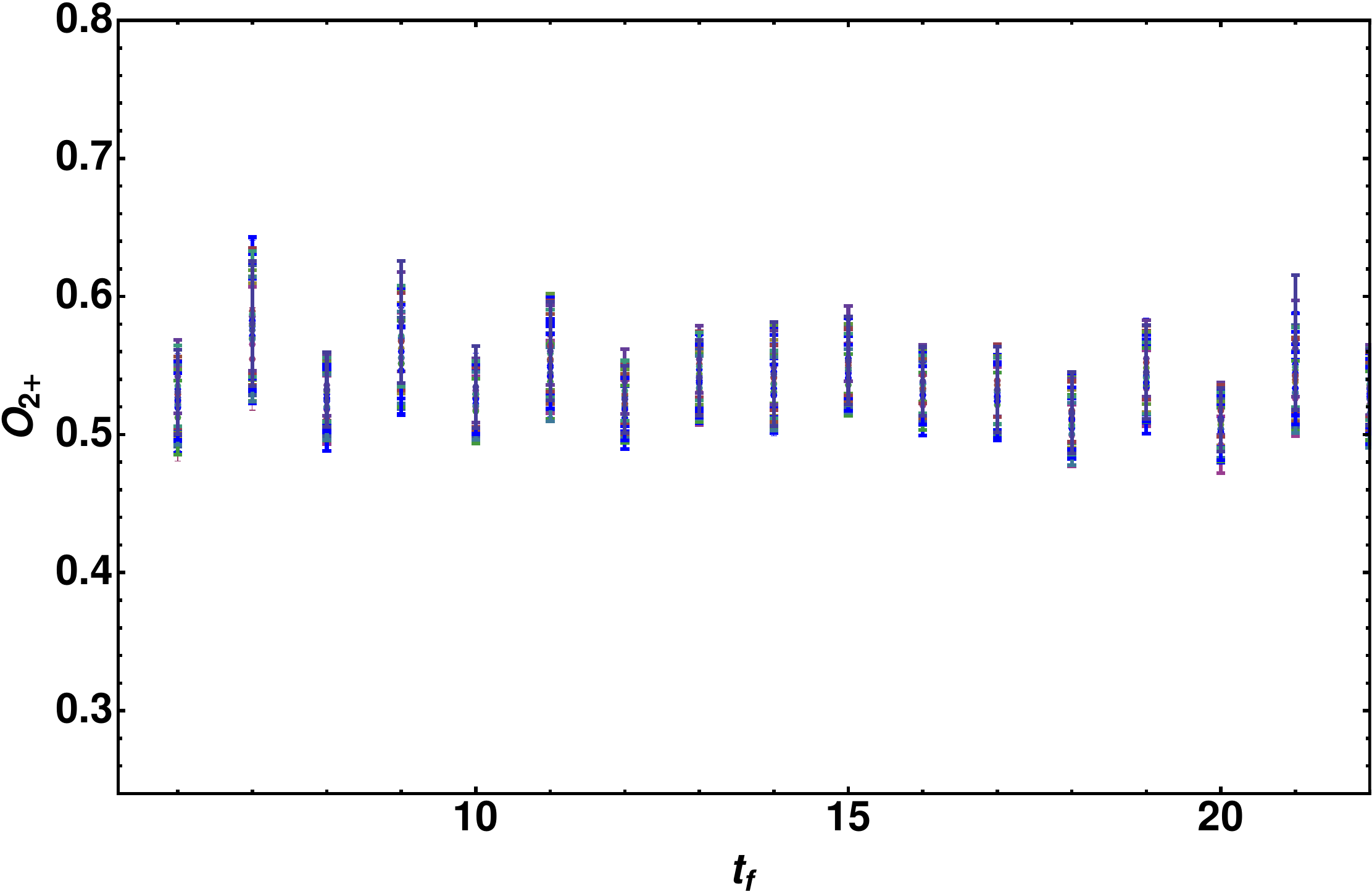}
\end{center}
\caption{\label{fig:EffMass}Plots of the ratio, $\calR$ (\Eq{ratio}), for a representative operator $\mathcal{O}_{2+}^{++}$, on the $m_{\pi} \sim 135$ MeV, $a$=0.12 fm, $L=48$ ensemble. On the left we show the ratio versus both times, $t_i,t_f$, while on the right we show only the plateau region versus $t_f$, with different values of $t_i$ represented by (overlapping) points.}
\end{figure}

In \Fig{EffMass}, we show representative plots of the ratio:
\beq
\label{eq:ratio}
\calR \equiv C_{3\mathrm{pt}}(t_i,t_f)/\left(C_{\pi}(t_i)C_{\pi}(t_f)\right) \ ,
\eeq 
where $C_{3\mathrm{pt}}$ is the three-point function sketched in \Fig{Ops}, and $C_{\pi}$ is a pion correlator, for $\mathcal{O}_{2+}^{++}$ on the physical pion mass, $L=48$, $a=0.12$ fm ensemble. For large $t_i$, $t_f$, this ratio approaches a constant corresponding to the desired matrix element. We plot the same data in two ways, a 3-dimensional plot of the effective mass as a function of the initial and final times, $t_i$ and $t_f$, respectively, as well as a traditional effective mass plotted versus $t_f$, where different-colored data points represent different values of $t_i$, for only the plateau region. This entire collection of data points may be fit to extract the matrix element. 

\begin{figure}
\begin{center}
\includegraphics[width=0.47\linewidth]{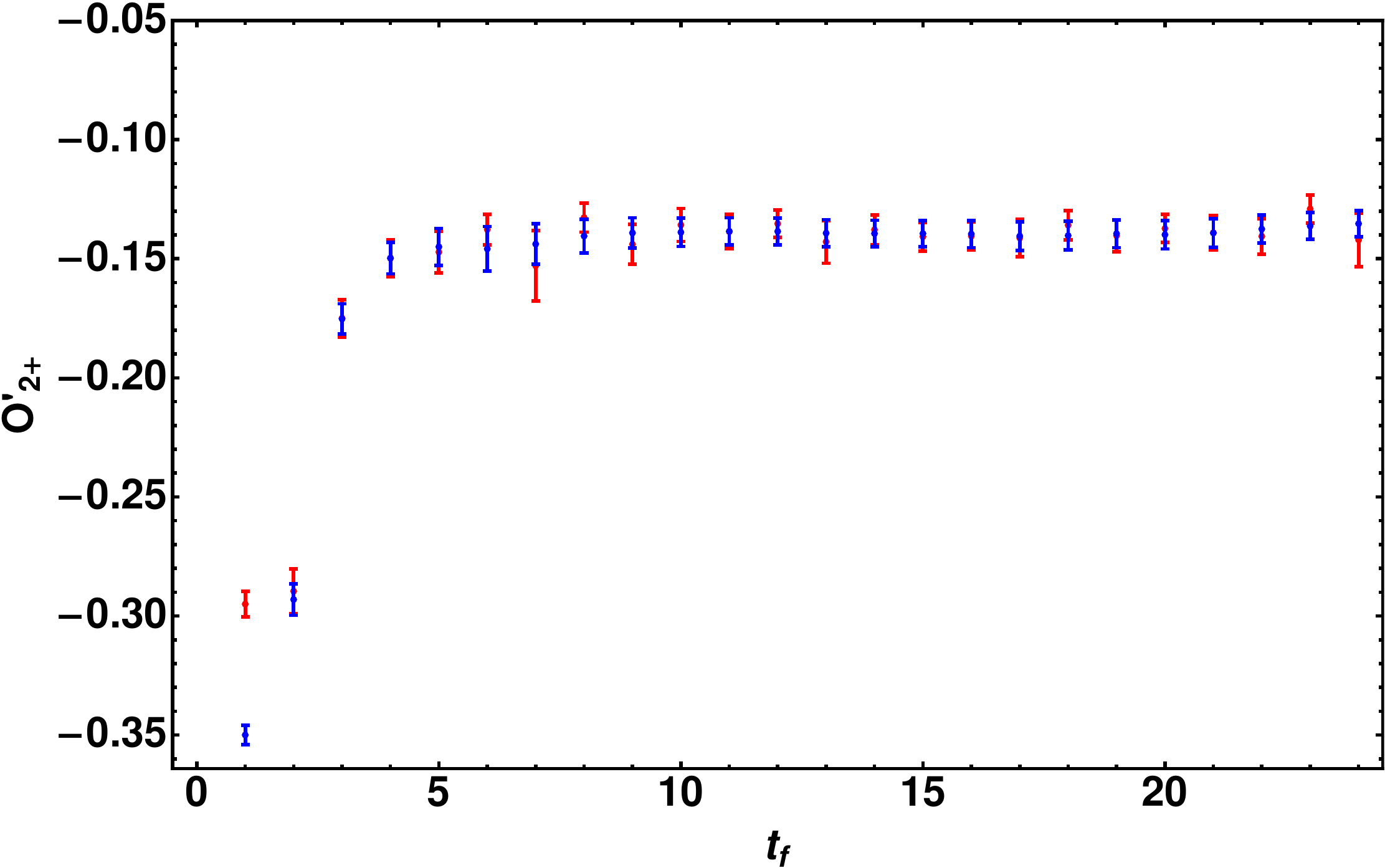}
\end{center}
\caption{\label{fig:pointvwall}The ratio, $\calR$ (\Eq{ratio}), versus $t_f$ for the color-mixed operator $\mathcal{O}_{2+}^{'++}$ on the $m_{\pi} \sim 135$ MeV, $a$=0.12 fm, $L=48$ ensemble, showing two types of interpolating fields for the pion: wall (red), and point (blue).}
\end{figure}

We find excellent signals on nearly all ensembles, requiring only a simple fit to a constant. This is likely due to the fact that in the ratio defined above, \Eq{ratio}, the contribution from the lowest thermal pion state is eliminated, which we find to be the leading contamination to the pion correlation function within the relevant time range. We also find little variation of the ratio using either our wall or point sources, as shown in \Fig{pointvwall}. This gives us additional confidence that excited state contamination is negligible within the time range plotted in \Fig{EffMass}.

\begin{figure}
\begin{center}
\includegraphics[width=0.47\linewidth]{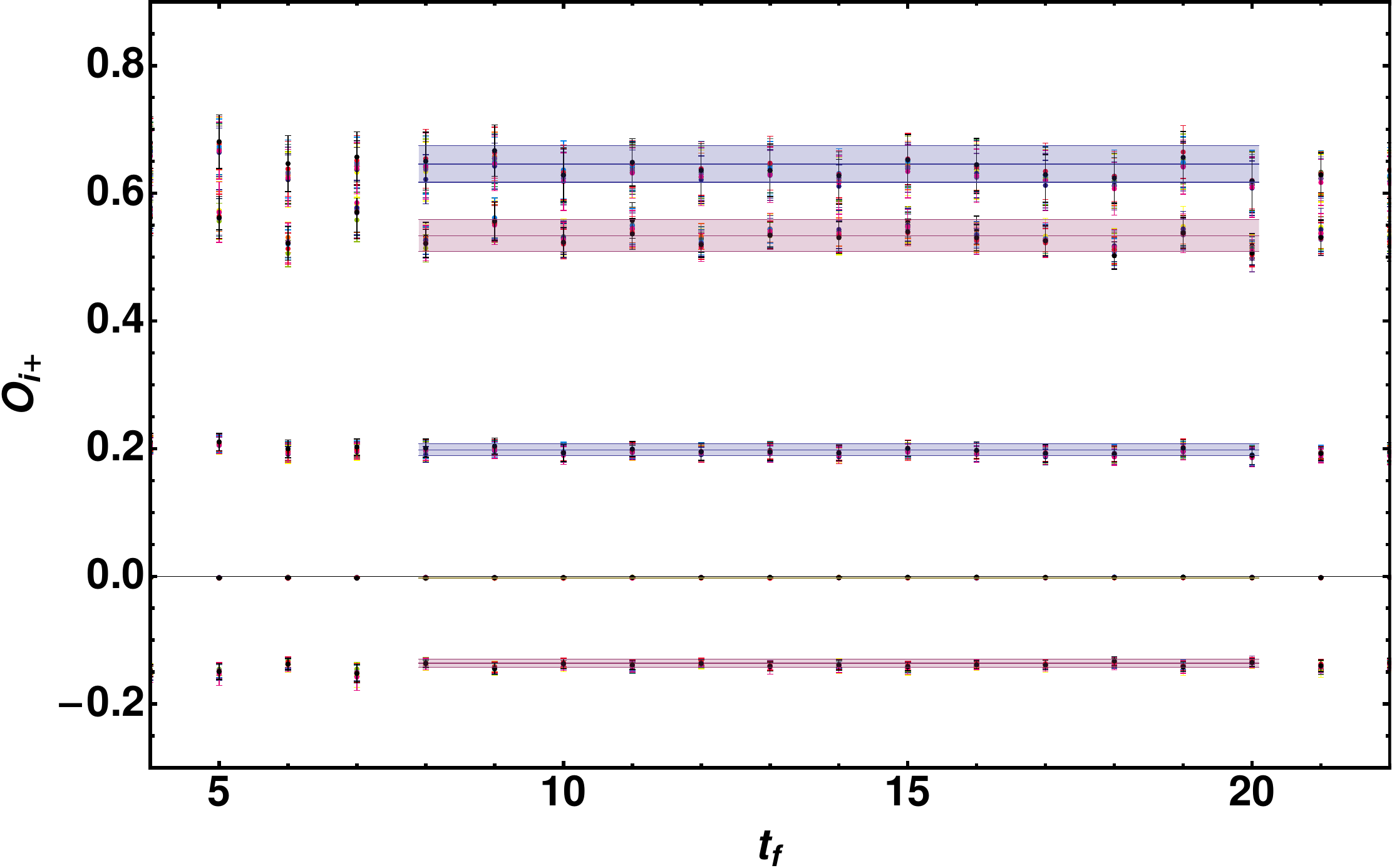}
\includegraphics[width=0.47\linewidth]{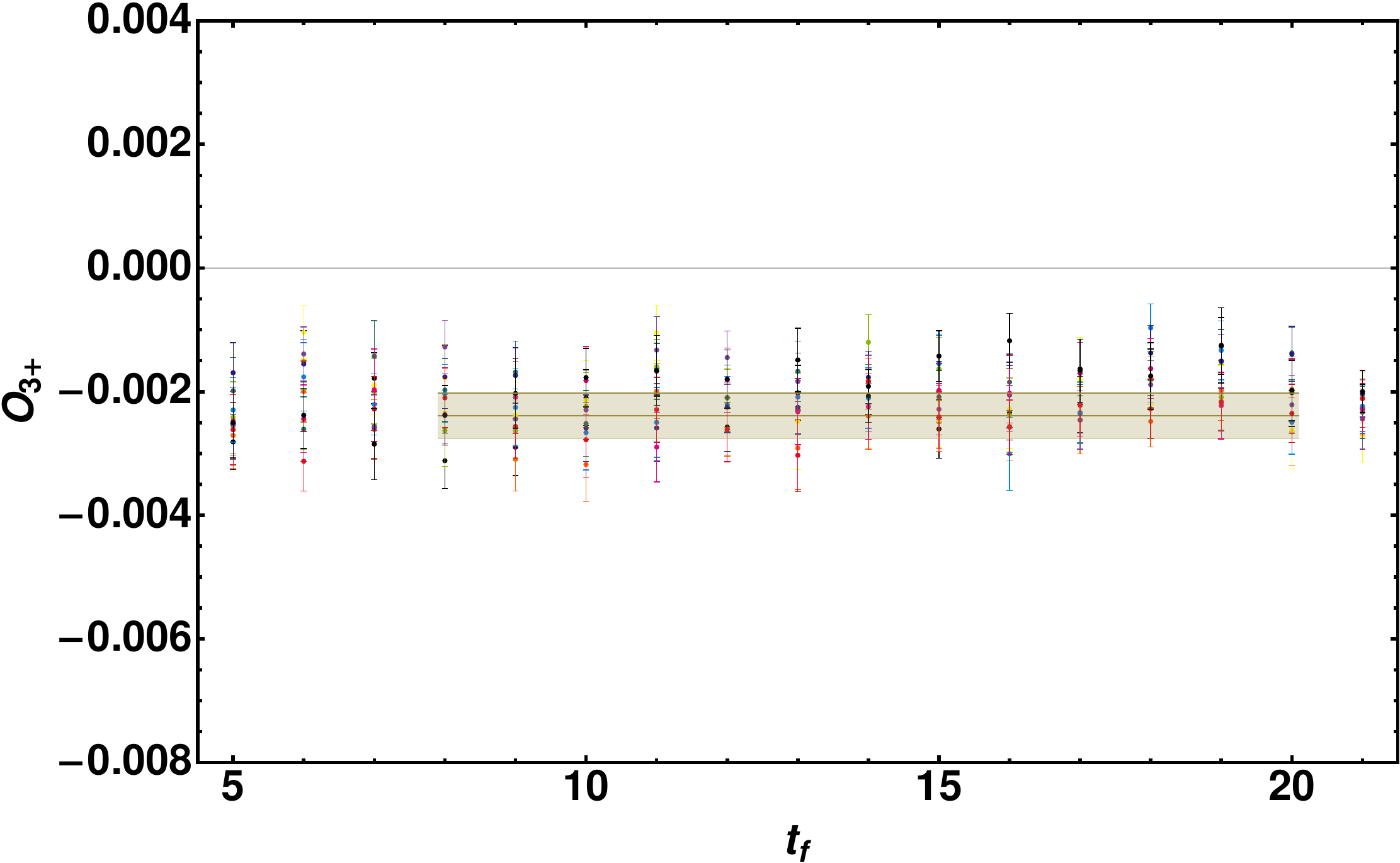}
\end{center}
\caption{\label{fig:EffMassfits}Ratio $\calR$ (\Eq{ratio}) versus $t_f$ for the five operators defined in \Eq{Ops} on the $m_{\pi} \sim 135$ MeV, $a$=0.12 fm, $L=48$ ensemble. Preliminary constant fits to each operator are shown as bands representing combined statistical and fitting systematic uncertainties. Operators which mix under renormalization are color-coded: $\mathcal{O}_{1+}^{++}, \mathcal{O}_{1+}^{'++}$: blue, $\mathcal{O}_{2+}^{++}, \mathcal{O}_{2+}^{'++}$: pink, $\mathcal{O}_{3+}^{++}$: yellow. On the right we zoom in to show $\mathcal{O}_{3+}^{++}$ in more detail.}
\end{figure}

In \Fig{EffMassfits} we plot preliminary fit results for the same ensemble for all five operators as horizontal bands representing combined statistical and fitting systematic errors. Matrix elements which mix under renormalization are shown with the same color. Due to the small sample size of our current data, a greatly reduced number of time slices are fit such that the correlation matrix is well-behaved. A larger number of sources will not only improve the statistics, but will also allow us to fit a larger set of points, greatly reducing the error bars on the fits. On the right we have plotted $\mathcal{O}_{3+}^{++}$ on a larger scale. As predicted by chiral counting, this operator is roughly two orders of magnitude smaller than the others. Even with our small sample size, we are able to cleanly resolve this operator.

\begin{figure}
\begin{center}
\includegraphics[width=0.47\linewidth]{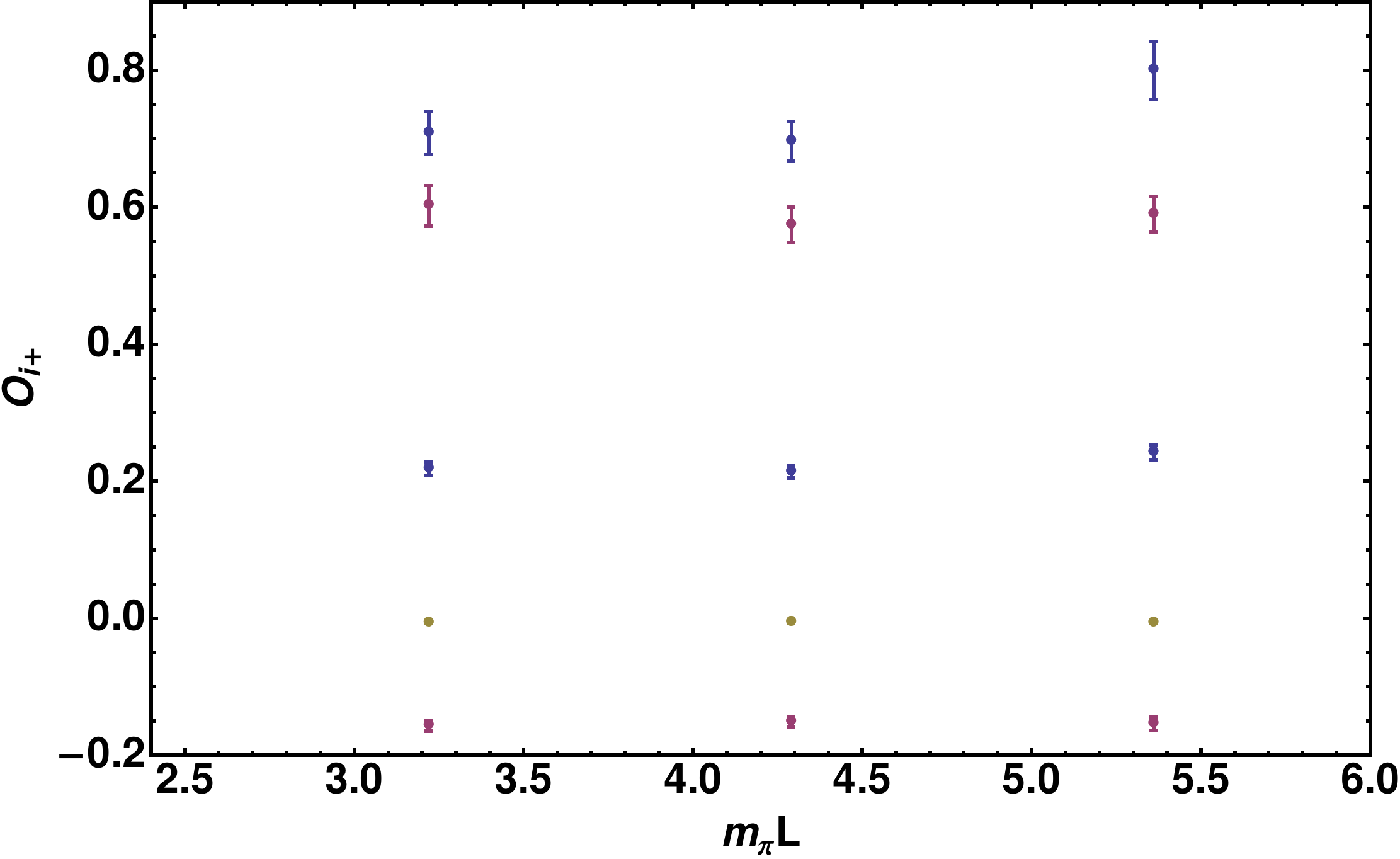}
\includegraphics[width=0.47\linewidth]{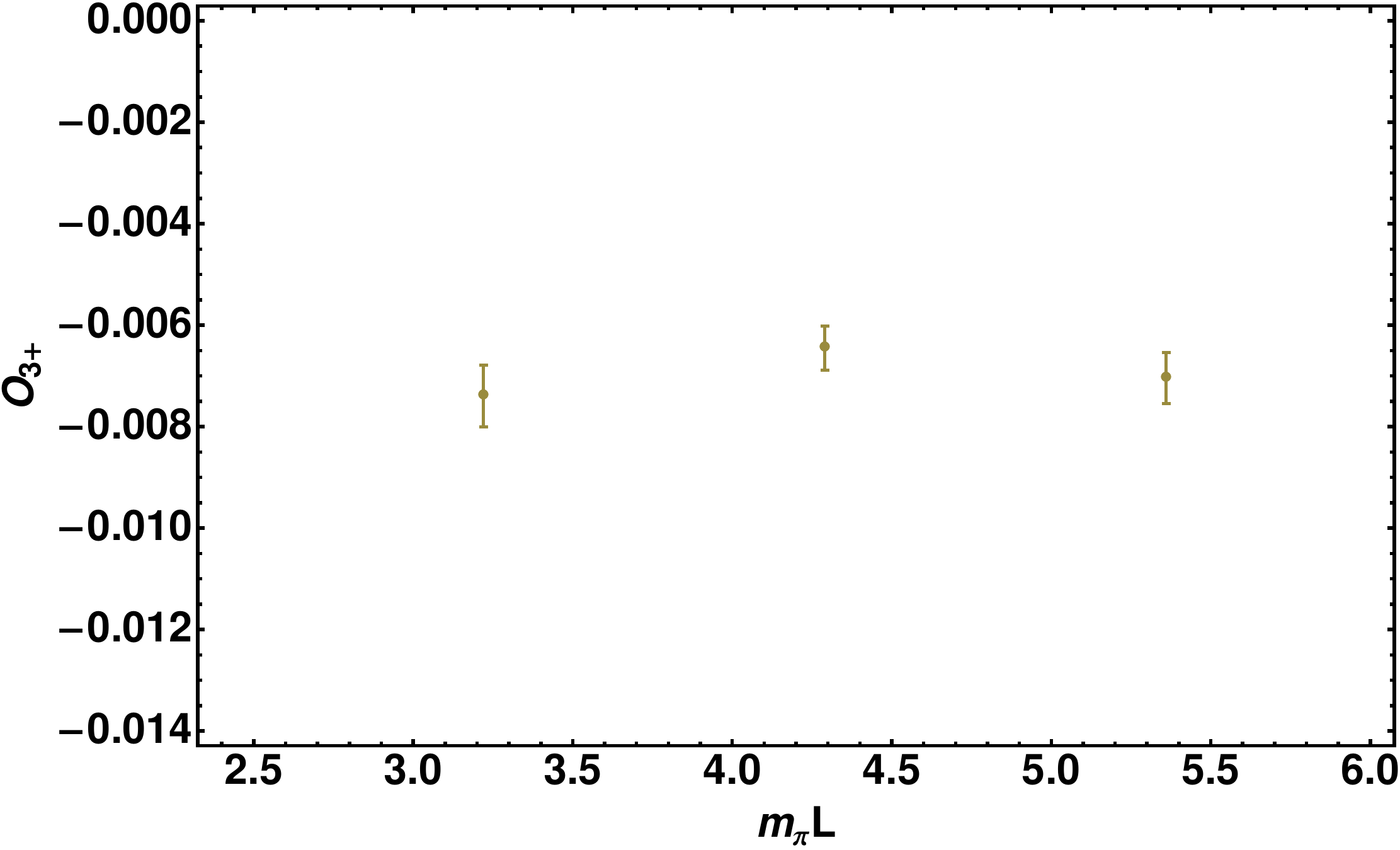}
\end{center}
\caption{\label{fig:volStudy}Results for all operators calculated on the $m_{\pi}\sim$220 MeV, $a$=0.12 fm ensembles versus $m_{\pi}L$. Color-coding is as defined in \Fig{EffMassfits}. On the right we zoom in to show $\mathcal{O}_{3+}^{++}$ in more detail.}
\end{figure}

\begin{figure}
\begin{center}
\includegraphics[width=0.47\linewidth]{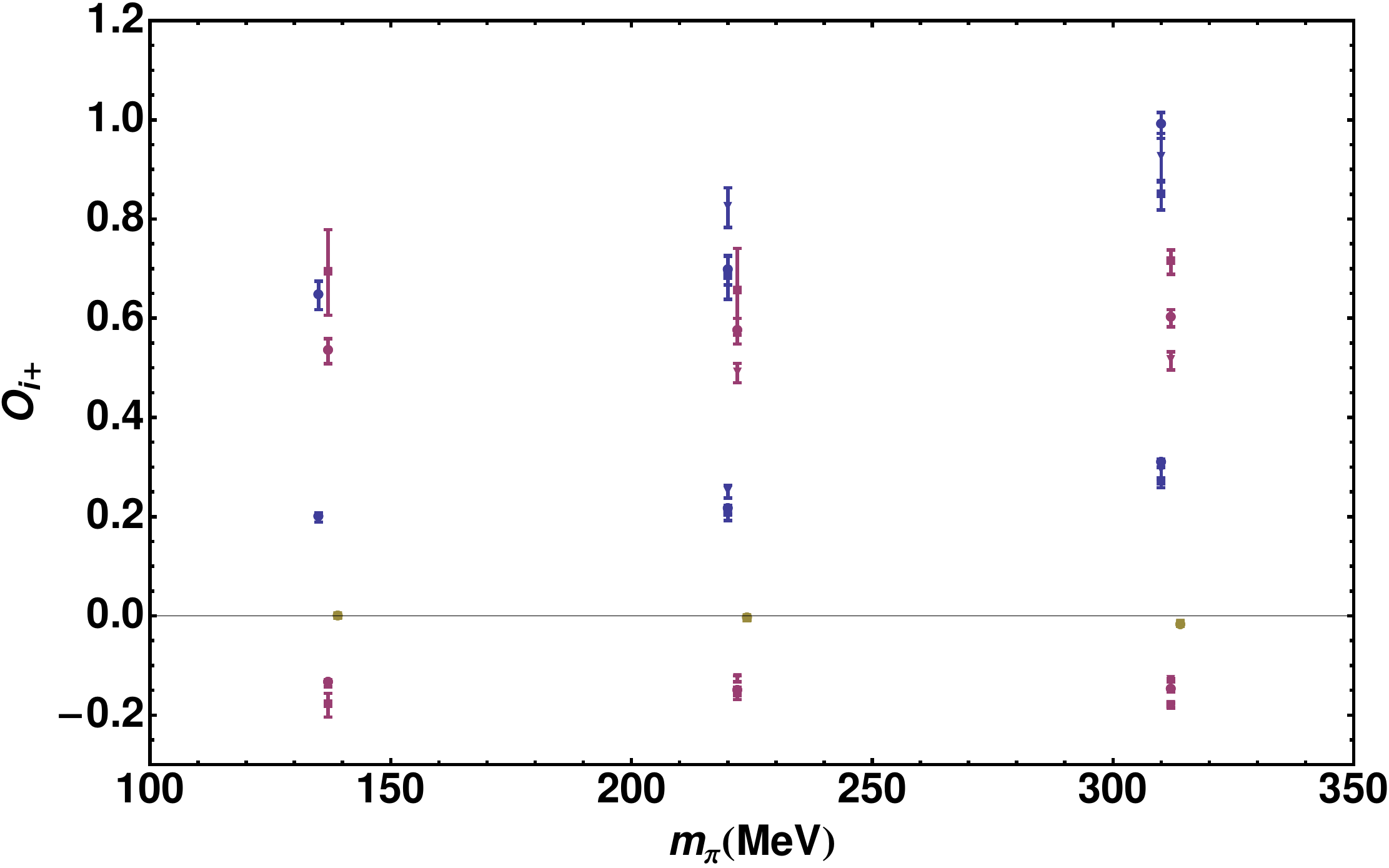}
\includegraphics[width=0.47\linewidth]{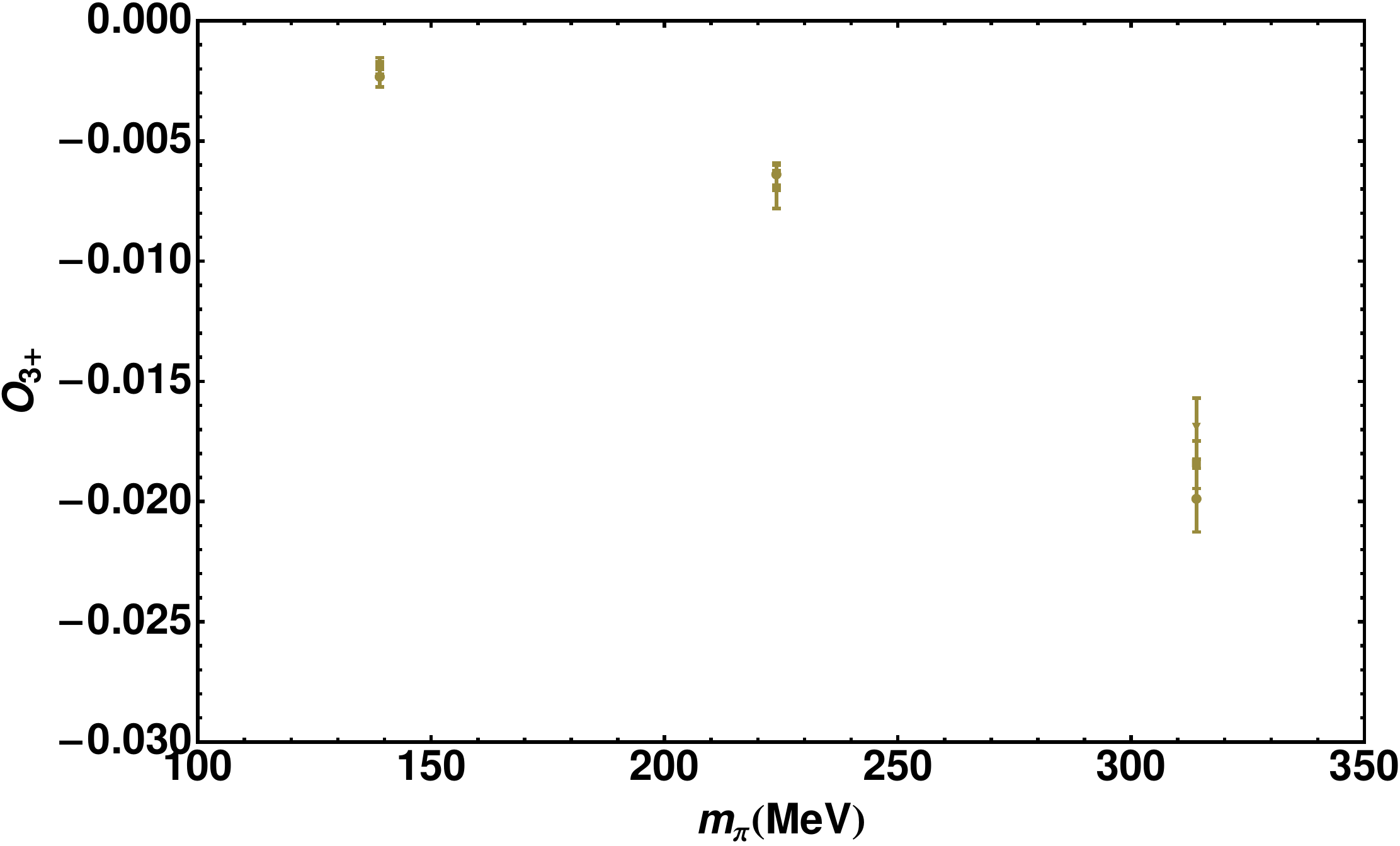}
\end{center}
\caption{\label{fig:vmpi}Results for all operators calculated on all ensembles in \Tab{ens} versus $m_{\pi}$. Points for different operators have been slightly displaced on the horizontal axis for ease of viewing. Color-coding is as defined in \Fig{EffMassfits}. On the right we zoom in to show $\mathcal{O}_{3+}^{++}$ in more detail.}
\end{figure}

\Fig{volStudy} shows a study of the finite volume effects of all operators, on the $m_{\pi} \sim 220$ MeV, $a=0.12$ fm ensembles. We find no significant variation of the results on these ensembles, therefore, finite volume effects appear to be negligible. In \Fig{vmpi} we plot the results for the operators calculated on all ensembles as a function of $m_{\pi}$. Due to the consistency found in the finite volume study, it is likely that the variation in the operators between different ensembles is caused by the different lattice spacings. We have not yet performed renormalization of these operators, including calculation of the operator mixing, so it is difficult to predict how large discretization effects will be. 

\begin{figure}
\begin{center}
\includegraphics[width=0.47\linewidth]{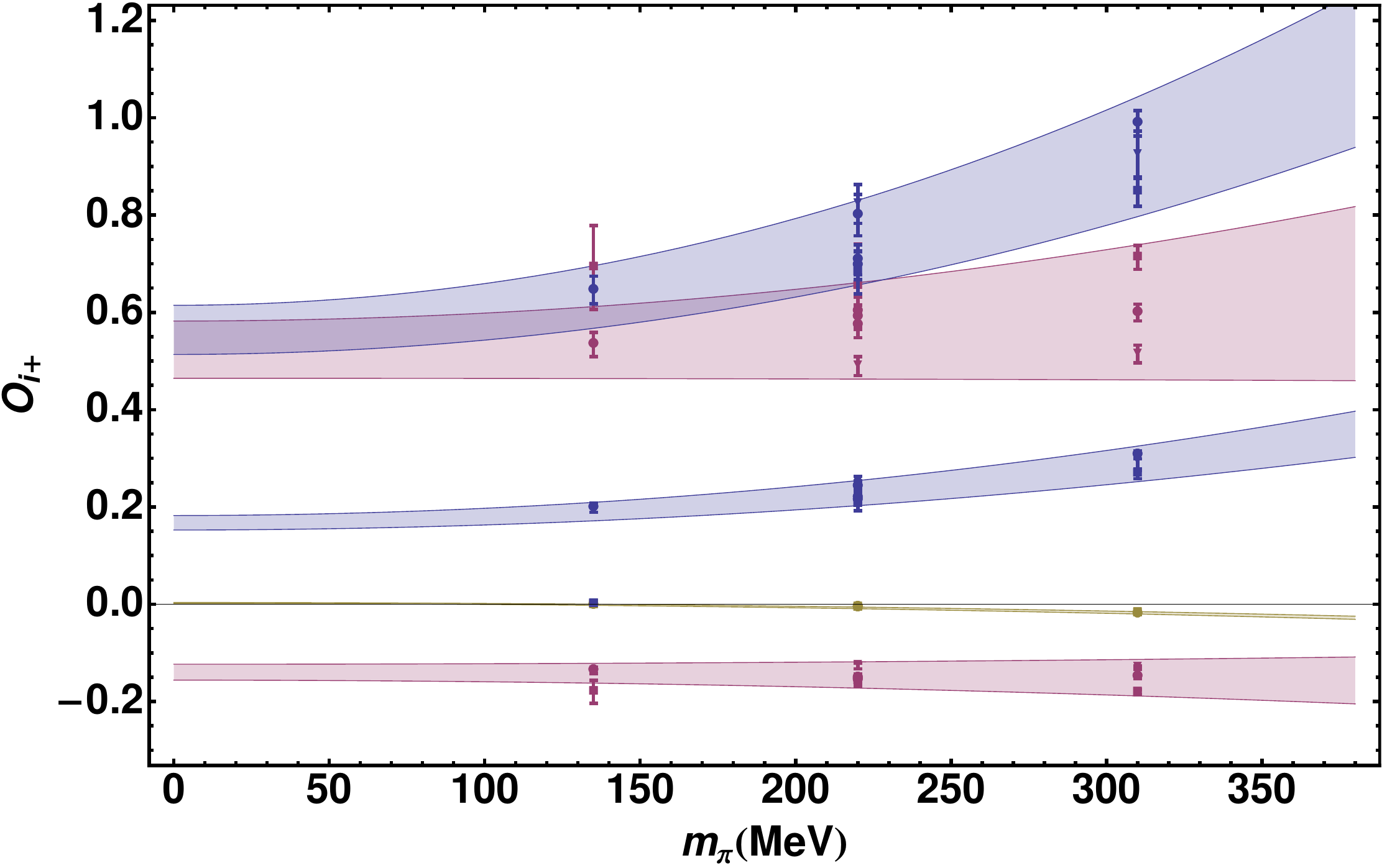}
\includegraphics[width=0.47\linewidth]{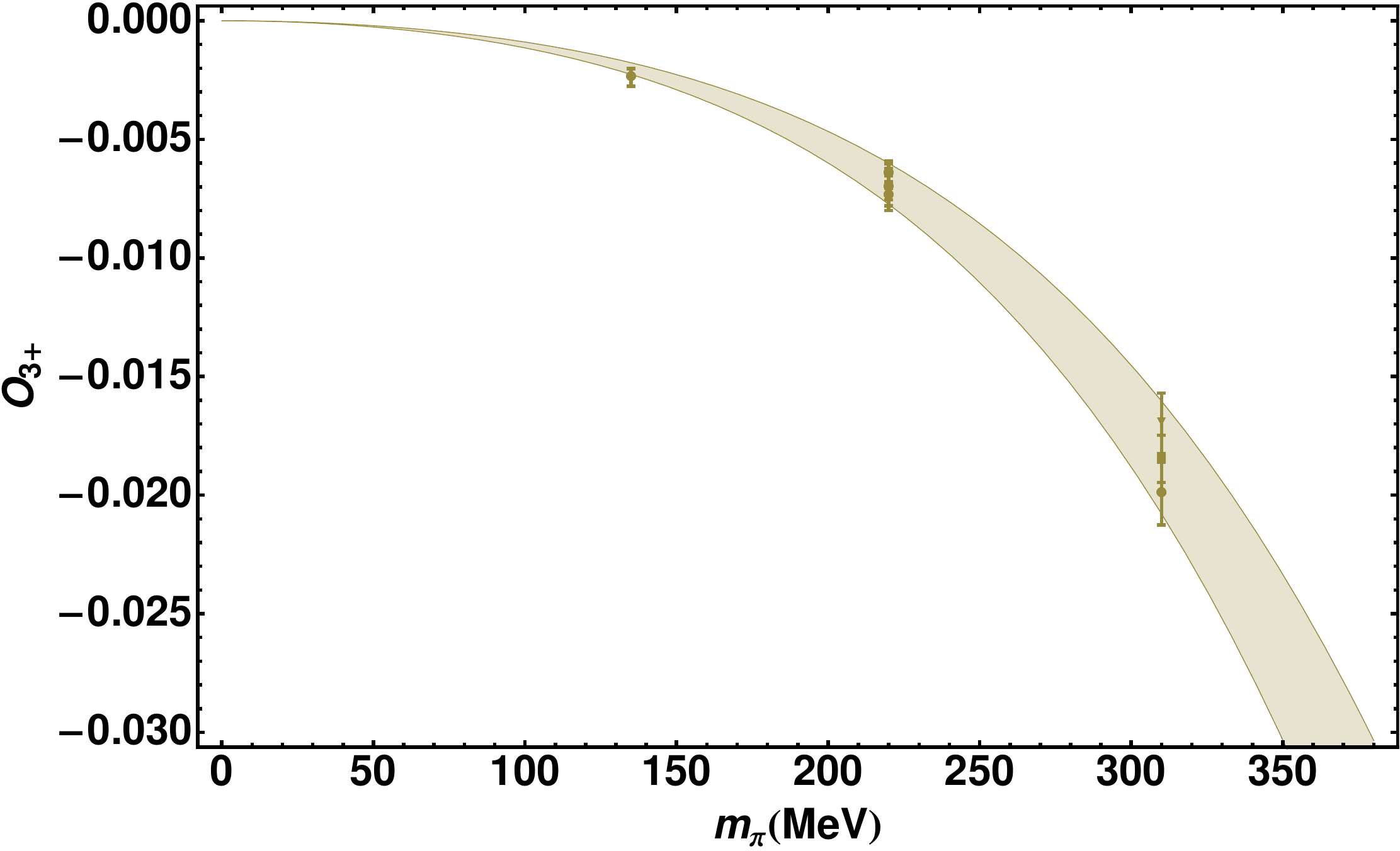}
\end{center}
\caption{\label{fig:mpiFits}Data from \Fig{vmpi}, along with polynomial fits to $m_{\pi}$ shown as shaded bands representing combined statistical and fitting systematic errors. $\mathcal{O}_{1+}^{++}, \mathcal{O}_{1+}^{'++}, \mathcal{O}_{2+}^{++},$ and $\mathcal{O}_{2+}^{'++}$ are fit to the form, $a+b m_{\pi}^2$, where $a$ and $b$ are fit parameters. On the right we zoom in on $\mathcal{O}_{3+}^{++}$, which is fit to the form $a m_{\pi}^2 + b m_{\pi}^4$.}
\end{figure}

We have performed simple fits to polynomials in $m_{\pi}$, and find that the pion mass dependence of the operators is roughly consistent with that expected from chiral effective theory. This is shown in \Fig{mpiFits}. Note, in particular, that $\mathcal{O}_{3+}^{++}$ fits well to a quadratic plus quartic fit in $m_{\pi}$, in agreement with the observation that there is no leading order contribution to this operator in chiral effective theory. Finally, we find that the shapes of the fit curves are similar for operators which mix under renormalization.

\section{Additional operators}

Now we will briefly discuss the calculation of the additional operators shown in \Fig{EFTOps}. At next-to-leading order we have $n \to p \pi$-type vertices. These types of vertices involve disconnected diagrams, requiring the computation of all-to-all propagators. Fortunately, most experimental efforts are focused on $0^+ \to 0^+$ nuclear transitions, where these vertices vanish due to parity. Thus, we likely do not need to consider these operators. 

\begin{figure}
\begin{center}
\includegraphics[width=0.45\linewidth]{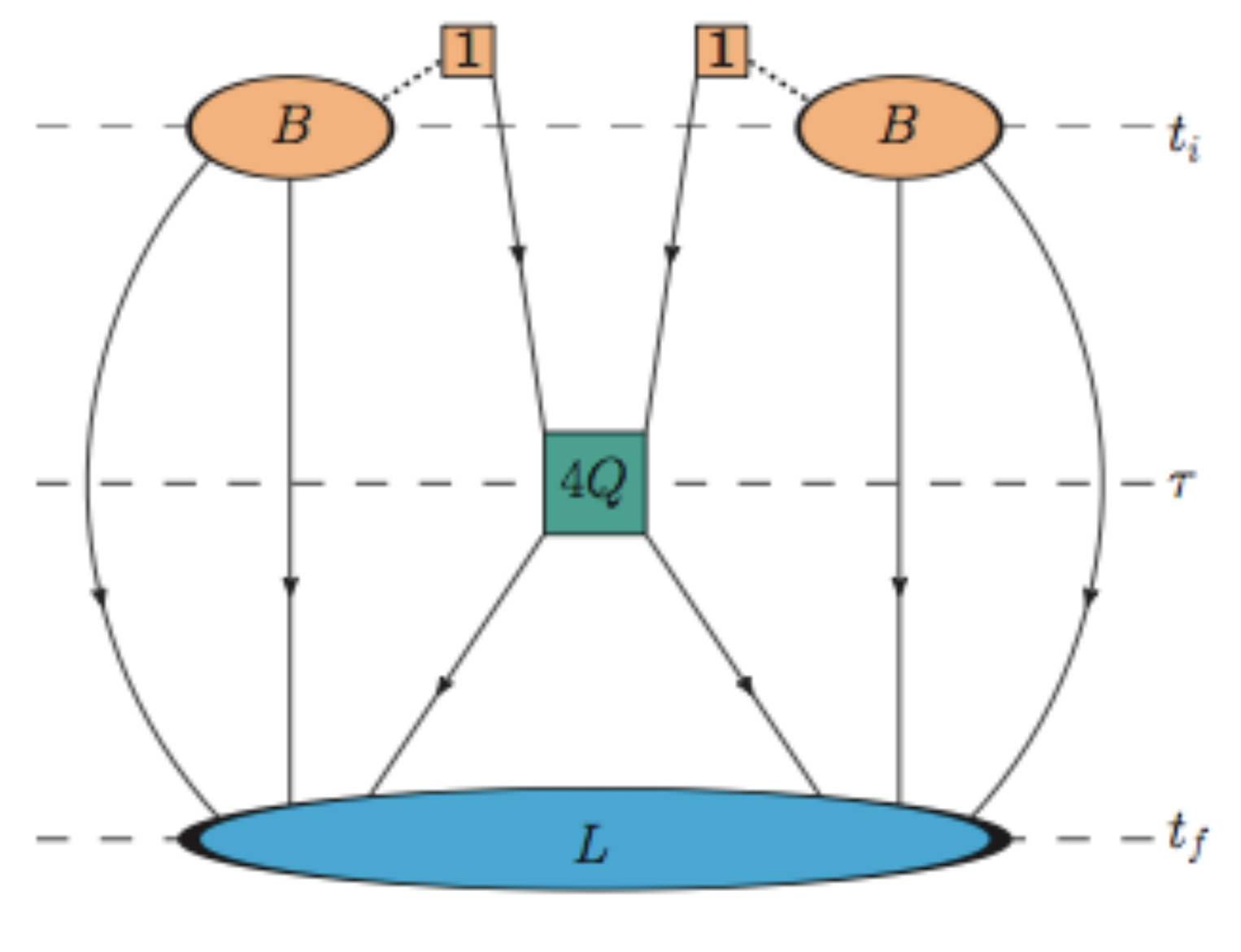}
\end{center}
\caption{\label{fig:NN}Sketch of the contraction contraction scheme used for two-nucleon contact operators, \Fig{EFTOps} (right). Two baryon blocks, $B$, are created at the source, contracted with the four-quark operator at time $\tau$ (quark lines extending from $B$ and $4Q$ are contracted via an identity matrix, $1$), then fully contracted at the sink with a tensor, $L$. Figure from \cite{Kurth:2015cvl}.}
\end{figure}

Finally, we shall discuss the two-nucleon contact operators (\Fig{EFTOps}, right). The setup for such a calculation is identical to that presented for $\Delta I =2$ nuclear parity violation \cite{Kurth:2015cvl}, and is sketched in \Fig{NN}. Two baryon blocks are created at $t_i$, tied up with the four-quark operator at $\tau$, and contracted using a single tensor, $L$, at $t_f$, in an extension of the unified contraction algorithm \cite{Doi:2012xd,Detmold:2012eu,Gunther:2013xj}. The two baryon blocks need to be projected onto the appropriate cubic irrep, $A_1^+$. The finite volume formalism for relating $2\to2$ matrix elements calculated from the lattice to the infinite volume result has been worked out in \cite{Briceno:2015tza}.

\begin{figure}
\begin{center}
\includegraphics[width=0.47\linewidth]{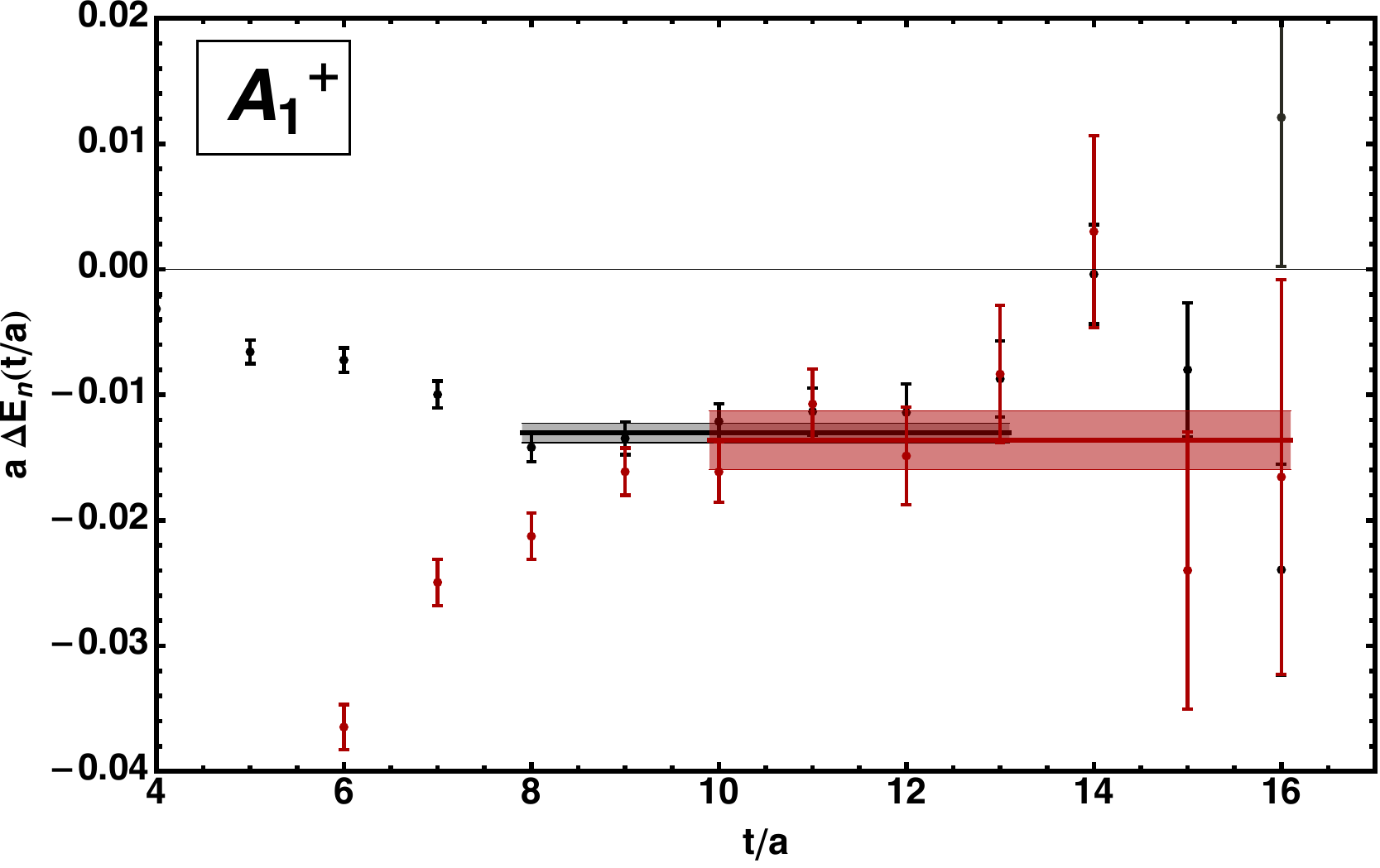}
\end{center}
\caption{\label{fig:A1p}Effective mass plots for two nucleons in the $A_1^+$ cubic irrep using different interpolating operators: spatially displaced two-nucleon operator at source, zero momentum projected two-nucleon operator at sink (black), and spatially displaced two-nucleon operators at both source and sink (red). The colored bands represent fits to the ground state for each set of data, including combined statistical and fitting systematic errors. Figure from \cite{Kurth:2015cvl}.}
\end{figure}

A major difference in this type of setup is that not all quark propagators are tied up with the four-quark operator. Therefore, we may only project onto definite momentum at one time if we wish to avoid the calculation of all-to-all propagators. We project the operator onto zero momentum transfer, leaving the total momentum of the system unspecified. For large Euclidean time, only the lowest possible total momentum will contribute. However, this setup requires us to compose two-nucleon operators in position space. We have performed a study of position space, two-nucleon operators for the $A_1^+$ irrep in \cite{Kurth:2015cvl} by comparing the effective mass plots for various two-nucleon configurations in position space with the known result calculated using momentum space sinks. Local operators, in which the two nucleons are created at the same spacetime point, offer the cheapest computational solution, however, we find very poor overlap with the ground state using these operators. Maximally displaced operators, in which the two nucleons are positioned at a distance $L/2$ from each other, seem to provide the best overlap with the ground state of the system (\Fig{A1p}). Results for the two-nucleon contact operators relevant for $\zeronubb$ will be presented in future work.

\section{Conclusions}

At present, there is an enormous experimental effort to detect $\zeronubb$, both to confirm the Majorana nature of neutrinos and measure their absolute mass scale, as well as to search for a source of lepton number violation which might contribute to the matter/antimatter asymmetry of the universe. In order to make better estimates of the expected lifetime of this process, as well as to learn about the particular mechanism underlying the $\zeronubb$, it is important to understand the contributions from both long- and short-range operators. In this work, we have shown preliminary lattice QCD results for the leading four-quark operators arising in a chiral effective theory framework on a variety of ensembles, and have performed a preliminary investigation of their pion mass dependencies. 

To connect the matrix element calculated from the lattice at the QCD scale to the relevant electroweak scale, renormalization of the operator must be performed. The renormalization for these operators is known perturbatively to two loops in the $\overline{\mathrm{MS}}$ scheme. We will perform non-perturbative renormalization to match on to this scheme in a forthcoming publication \cite{zeronubb}, as well as perform extrapolations to the physical pion mass and the continuum.

\acknowledgments
Numerical calculations were performed with the \texttt{Chroma} software suite \cite{Edwards:2004sx} with \texttt{QUDA} solvers \cite{Clark:2009wm,Babich:2011np} on Surface at LLNL, supported by the LLNL Multiprogrammatic and Institutional Computing program through a Tier 1 Grand Challenge award, and on Titan, a resource of the Oak Ridge Leadership Computing Facility at the Oak Ridge National Laboratory, which is supported by the Office of Science of the U.S. Department of Energy under Contract No. DE-AC05-00OR22725, through a 2016 INCITE award. This work was performed under the auspices of the U.S. Department of Energy by Lawrence Livermore National Laboratory under contract~{DE-AC52-07NA27344}.
The work of A.N. was supported in part by the U.S. Department of Energy under grant DE-SC00046548. 
The work of C.C.C. was supported in part by the U.S. Department of Energy, Office of Science, Office of Nuclear Physics under the DOE Early Career Research Program under Award Number NQCDAWL.
The work of T.K. was supported in part by the Director, Office of Science, of the U.S. Department of Energy under Contract No. DE-AC02-05CH11231.
The work of A.W-L. was supported in part by the U.S. Department of Energy, Office of Science: 
Office of Advanced Scientific Computing Research, Scientific Discovery through Advanced Computing (SciDAC) program under Award Number KB0301052;
the Office of Nuclear Physics under Contract number DE-AC02-05CH11231 and the DOE Early Career Research Program under Award Number NQCDAWL.

\bibliographystyle{physrev} %%% physical review
\bibliography{NN} %%% bib file

\begin{thebibliography}{10}

\bibitem{Haxton:1985am}
W.~C. Haxton and G.~J. Stephenson,
\newblock Prog. Part. Nucl. Phys. {\bf 12}, 409 (1984).

\bibitem{Geesaman:2015fha}
A.~Aprahamian {\em et~al.},
\newblock (2015).

\bibitem{Pascoli:2006ci}
S.~Pascoli, S.~T. Petcov, and A.~Riotto,
\newblock Nucl. Phys. {\bf B774}, 1 (2007), hep-ph/0611338.

\bibitem{Davidson:2008bu}
S.~Davidson, E.~Nardi, and Y.~Nir,
\newblock Phys. Rept. {\bf 466}, 105 (2008), 0802.2962.

\bibitem{Buchmuller:2005eh}
W.~Buchmuller, R.~D. Peccei, and T.~Yanagida,
\newblock Ann. Rev. Nucl. Part. Sci. {\bf 55}, 311 (2005), hep-ph/0502169.

\bibitem{Dell'Oro:2016dbc}
S.~Dell'Oro, S.~Marcocci, M.~Viel, and F.~Vissani,
\newblock Adv. High Energy Phys. {\bf 2016}, 2162659 (2016), 1601.07512.

\bibitem{Bilenky:2012qi}
S.~M. Bilenky and C.~Giunti,
\newblock Mod. Phys. Lett. {\bf A27}, 1230015 (2012), 1203.5250.

\bibitem{schechter1982neutrinoless}
J.~Schechter and J.~W. Valle,
\newblock Physical Review D {\bf 25}, 2951 (1982).

\bibitem{Nieves:1984sn}
J.~F. Nieves,
\newblock Phys. Lett. {\bf B147}, 375 (1984).

\bibitem{Takasugi:1984xr}
E.~Takasugi,
\newblock Phys. Lett. {\bf B149}, 372 (1984).

\bibitem{Rosen:1992qa}
S.~P. Rosen,
\newblock {Double beta decay},
\newblock in {\em {The Benjamin Franklin Symposium in Celebration of the
  Discovery of the Neutrino Philadelphia, Pennsylvania, April 29-May 1, 1992}},
  pp. 31--48, 1992, hep-ph/9210202.

\bibitem{Hirsch:2006yk}
M.~Hirsch, S.~Kovalenko, and I.~Schmidt,
\newblock Phys. Lett. {\bf B642}, 106 (2006), hep-ph/0608207.

\bibitem{Prezeau:2003xn}
G.~Prezeau, M.~Ramsey-Musolf, and P.~Vogel,
\newblock Phys. Rev. {\bf D68}, 034016 (2003), hep-ph/0303205.

\bibitem{Takahashi:2012}
Y.~Takahashi,
\newblock {The Fierz Identities},
\newblock in {\em {Progress in Quantum Field Theory}}, edited by H.~Ezawa and
  S.~Kamefuchi, p. 121, North-Holland, Amsterdam, 1986.

\bibitem{Graesser:2016bpz}
M.~L. Graesser,
\newblock (2016), 1606.04549.

\bibitem{Buras:2000if}
A.~J. Buras, M.~Misiak, and J.~Urban,
\newblock Nucl. Phys. {\bf B586}, 397 (2000), hep-ph/0005183.

\bibitem{Savage:1998yh}
M.~J. Savage,
\newblock Phys. Rev. {\bf C59}, 2293 (1999), nucl-th/9811087.

\bibitem{Aoki:2016frl}
S.~Aoki {\em et~al.},
\newblock (2016), 1607.00299.

\bibitem{Buchoff:2012bm}
M.~I. Buchoff, C.~Schroeder, and J.~Wasem,
\newblock PoS {\bf LATTICE2012}, 128 (2012), 1207.3832.

\bibitem{Bazavov:2012xda}
MILC, A.~Bazavov {\em et~al.},
\newblock Phys. Rev. {\bf D87}, 054505 (2013), 1212.4768.

\bibitem{Bazavov:2015yea}
MILC, A.~Bazavov {\em et~al.},
\newblock Phys. Rev. {\bf D93}, 094510 (2016), 1503.02769.

\bibitem{Brower:2012vk}
R.~C. Brower, H.~Neff, and K.~Orginos,
\newblock (2012), 1206.5214.

\bibitem{Action}
E.~Berkowitz {\em et~al.},
\newblock {in preparation}.

\bibitem{Clark:2009wm}
M.~A. Clark, R.~Babich, K.~Barros, R.~C. Brower, and C.~Rebbi,
\newblock Comput. Phys. Commun. {\bf 181}, 1517 (2010), 0911.3191.

\bibitem{Babich:2011np}
R.~Babich {\em et~al.},
\newblock {Scaling Lattice QCD beyond 100 GPUs},
\newblock in {\em {SC11 International Conference for High Performance
  Computing, Networking, Storage and Analysis Seattle, Washington, November
  12-18, 2011}}, 2011, 1109.2935.

\bibitem{Luscher:2013cpa}
M.~Luscher,
\newblock JHEP {\bf 04}, 123 (2013), 1302.5246.

\bibitem{Luscher:2011bx}
M.~Luscher and P.~Weisz,
\newblock JHEP {\bf 02}, 051 (2011), 1101.0963.

\bibitem{Luscher:2010iy}
M.~Lüscher,
\newblock JHEP {\bf 08}, 071 (2010), 1006.4518,
\newblock [Erratum: JHEP03,092(2014)].

\bibitem{Narayanan:2006rf}
R.~Narayanan and H.~Neuberger,
\newblock JHEP {\bf 03}, 064 (2006), hep-th/0601210.

\bibitem{GradFlow}
C.~Monahan and K.~Orginos,
\newblock {private communication}.

\bibitem{Kurth:2015cvl}
T.~Kurth {\em et~al.},
\newblock PoS {\bf LATTICE2015}, 329 (2016), 1511.02260.

\bibitem{Doi:2012xd}
T.~Doi and M.~G. Endres,
\newblock Comput. Phys. Commun. {\bf 184}, 117 (2013), 1205.0585.

\bibitem{Detmold:2012eu}
W.~Detmold and K.~Orginos,
\newblock Phys. Rev. {\bf D87}, 114512 (2013), 1207.1452.

\bibitem{Gunther:2013xj}
J.~Günther, B.~C. Toth, and L.~Varnhorst,
\newblock Phys. Rev. {\bf D87}, 094513 (2013), 1301.4895.

\bibitem{Briceno:2015tza}
R.~A. Briceño and M.~T. Hansen,
\newblock Phys. Rev. {\bf D94}, 013008 (2016), 1509.08507.

\bibitem{zeronubb}
E.~Berkowitz {\em et~al.},
\newblock {in preparation}.

\bibitem{Edwards:2004sx}
SciDAC, LHPC, UKQCD, R.~G. Edwards and B.~Joo,
\newblock Nucl. Phys. Proc. Suppl. {\bf 140}, 832 (2005), hep-lat/0409003,
\newblock [,832(2004)].

\end{thebibliography}

\end{document}